# Atomistic Insights into the Chain-Length-Dependent Antifreeze Activity of Oligoprolines

*Wentao Yang, Yucong Liao, Zhaoru Sun\**

School of Physical Science and Technology, ShanghaiTech University, Shanghai 201210, China

**ABSTRACT:** Oligoproline is a simple yet highly potent cryoprotectant, but the molecular basis underlying its nonmonotonic ice recrystallization inhibition (IRI) activity depending on the degree of polymerization (DP)—particularly the superior performance of DP=8 (P8) over longer (e.g., P15) oligomers—remains elusive. Using molecular dynamics simulations, we show that the IRI activity originates from the combined effect of single-molecule conformation and multi-molecule aggregation. P8 outperforms P15 primarily arises from its higher proportion of the random coil (C) conformation, which effectively enhances ice-binding ability and resistance to ice engulfment than the linear (L) conformation with perfect PPII helix. Moreover, at high concentrations (>40 mg/mL), P15 tends to form soluble amorphous aggregates, reducing its effective coverage on the ice surface and thereby further diminishing its IRI efficiency. These findings provide atomistic insight into the structure-activity relationship of oligoprolines and offer a framework for understanding similar nonmonotonic effect in other antifreeze polymers.

## 1. INTRODUCTION

Cryopreservation is a pivotal technique for the long-term preservation of cells, tissues, and organs in biotechnology and biomedicine,[1-3] providing invaluable support for scientific research[4-7] and clinical applications.[8-11] The main challenge in the freeze-thaw process of cryopreservation is uncontrolled ice recrystallization (IR), an Ostwald ripening process[12-14] in which larger ice crystals grow at the expense of smaller ones, leading to mechanical damage and osmotic shock, ultimately resulting in cell death.[15,16] Fortunately, numerous antifreeze materials have been shown to effectively control ice crystal growth through the widely accepted adsorption-inhibition mechanisms,[17-20] in which antifreeze agents bind to ice and prevent ice growth, thereby mitigating the detrimental effects of IR.[21-30] Among these materials, polymers exhibit considerable ice recrystallization inhibition (IRI) activity, broad availability, low cost, and tunable properties (e.g., degree of polymerization (DP), monomer species, etc.),[31-34] making them promising candidates for cryopreservation applications and having attracted tremendous research interest.[35-43] Nevertheless, the rational design and development of optimized IRI-active polymers for practical industrial and medical applications remains an ongoing challenge in this field.[44-46]

It is widely recognized that the IRI activity of polymers is directly correlated with their chain length or DP.[14,31,47,48] Typically, the IRI activity increases monotonically with chain length/DP, as demonstrated in a variety of materials including poly(vinyl alcohol) (PVA),[14] poly(L-alanine-co-L-lysine),[35] nanocellulose,[49] and others.[50-54] This phenomenon can be explained by a well-established mechanism that the longer chain provides more ice-binding sites, along with a larger effective volume and increased contact area with the ice surface. This enhances the ice-binding strength and prevents the polymer being engulfed by the advancing ice front, thereby improving IRI efficiency.[44,55]

However, certain polymers, such as oligoproline,[56] zwitterionic poly(carboxybetaine methacrylate),[57] thymine oligomer,[58] exhibit an intriguing yet poorly understood nonmonotonic relationship between IRI activity and chain length within a defined oligomeric range. In this case, the IRI activity initially increases but then declines as the chain length increases, with the highest activity observed at an intermediate chain length. Among these, oligoproline is the simplest and most representative system of this phenomenon, demonstrating excellent biocompatibility and significant potential for practical applications in cryopreservation compared to other polymers.[33,56] Experimental studies[56] have shown that oligoproline with DP=8 (P8) exhibits the highest activity, outperforming both shorter (i.e., P3) and longer (i.e., P15) chains. This suggests that the mechanism underlying the relationship between IRI activity and chain length is more complex in oligoproline than in conventional polymers such as PVA.[55] Uncovering this mechanism is essential to deepen our understanding of its structure-activity relationship and the nonmonotonic effect observed in other polymers, such as thymine oligomer.[58]

One interpretation of this nonmonotonic phenomenon is that oligoproline inhibits ice growth by binding to ice through its methylene groups within the linear amphipathic polyproline II (PPII) helix structure.[59] Preservation of this unique PPII helix dictates its IRI activity.[33] For example, Wang et al.[56] hypothesized that P8's superior activity is attributed to its perfect PPII helix structure, which facilitates its binding to the ice surface. In contrast, P3 is too short to form the PPII structure, and P15 primarily adopts the coiled conformation with a reduced PPII content. However, recent findings by Rojas et al.[60] challenge this view. Their circular dichroism (CD) experiments revealed that the PPII content of oligoproline increases with DP, with P8 exhibiting a more disordered structure compared to longer oligoprolines. This suggests that the PPII content alone cannot explain this nonmonotonic effect. Additionally, recent study[57] on another polymer system, zwitterionic poly(carboxybetaine methacrylate), suggested that the nonmonotonic effect originates from the degree of molecular extension, which directly influences the contact area and determines its IRI activity. Therefore, a comprehensive investigation of the molecular conformation of oligoproline and its impact on IRI activity at the atomic level is imperative for a complete understanding of the nonmonotonic effect.

In addition to the debates on single-molecule conformation, aggregation has also been proposed to influence the IRI activity of polymers.[61-66] In the simplest scenario, aggregation leads to the precipitation of insoluble aggregates, thereby reducing the effective concentration of antifreeze agents in solution and diminishing IRI activity, as observed in the high-molecular-weight tamarind seed polysaccharides.[47] However, when soluble aggregates are formed without noticeable precipitation, the impact of aggregation on IRI activity becomes more complex, as the effective concentration of antifreeze agents remains unchanged. Previous studies have shown that soluble aggregation (including assembly) can either enhance IRI activity (e.g., saffron molecules,[61] some self-assembled peptides[62]) or suppress it (e.g., isotactic PVA,[65] nanocelluloses[66,67]). Current evidence suggests that only the formation of ordered aggregates exposing ice-binding site-like structures—similar to those found in antifreeze proteins (AFPs)—can enhances the binding strength of antifreeze agents to ice, thereby increasing IRI activity.[63,64,68] Therefore, the influence of soluble aggregates on IRI activity is highly dependent on their microscopic morphology. Although experimental studies have reported that polyprolines (PPro) can form soluble aggregate at room temperature,[69-71] the microstructure of the aggregate and its impact on IRI activity remain unclear.

In this work, we employ all-atom molecular dynamics (MD) simulations to investigate the molecular mechanism underlying the nonmonotonic relationship between IRI activity and DP in oligoprolines (i.e., P8 > P3 and P8 > P15). Firstly, metadynamics simulations reveal that all these oligoprolines adopt two stable conformations—linear (L) and random coil (C)—both dominated by the PPII helix, with the total PPII content increases with DP, which aligns well with experimental measurements.[60] Further analyses demonstrate that the nonmonotonic IRI–DP relationship is governed by the C conformation content rather than the PPII helix structure. Specifically, P8's superior IRI activity compared to P15 primarily arises from its higher C conformation content, which facilitates stronger ice binding and greater resistance to ice engulfment. In contrast, despite its high C conformation content, P3's short length renders it susceptible to rapid engulfment by the advancing ice front, resulting in minimal IRI activity. Additionally, our results show that P15 exhibits a strong tendency to form soluble, amorphous aggregates at elevated concentrations (>40 mg/mL), limiting the effective ice surface coverage and further diminishing its IRI efficacy. This aggregation explains the experimental observation that difference in IRI activity between P8 and P15 becomes significantly pronounced at high concentrations (~50 mg/mL).[56] Collectively, our simulations provide a comprehensive molecular understanding of the chain-length-dependent IRI activity of oligoprolines.

## 2. METHODS

**2.1. Molecular dynamics simulations.** All molecular dynamics simulations were carried out by GROMACS 2020.7 packages[72] using the all-atomistic CHARMM36 forcefield[73] and the TIP4P/Ice water model.[74] The melting temperature of ice Ih in this water model is 270 K, in good agreement with the experimental value of 273.15 K.[75] The cutoffs for the van der Waals and Coulombic interactions were set to 1.2 nm, and long-range electrostatic interactions were evaluated with the particle-mesh Ewald (PME) algorithm.[76] The LINCS algorithm[77] was employed to constrain the covalent chemical bonds including hydrogen atoms. The equations of motion were integrated using the leapfrog method with a time step of 2 fs across all simulations. Periodic boundary conditions were applied in three directions. The temperature T and pressure P for production runs were controlled with the Nosé–Hoover thermostat[78,79] and Parrinello-Rahman barostat,[80] with time constants of 0.5 and 2.0 ps, respectively. The pressure was set to 1 atm in all NPT-MD simulations.

To investigate the ice growth inhibition activity of oligoprolines with DP=3, 8, and 15, the polymers in stable conformations (obtained from metadynamics simulations, see Section 2.2) were placed on top of four layers of hexagonal ice generated from the Genice program.[81] The system was then solvated with ~10000 water molecules, as previous experiments have reported that the oligoproline exhibits IRI activity at concentrations of 20 mg/mL.[33,56] The oligoproline molecules were positioned approximately 1.0 nm above the primary prismatic plane of the ice crystal (see Figure S1), maintaining consistent proline monomer concentrations across all DPs, using 5, 2, and 1 chains for DP = 3, 8, and 15, respectively. The initial dimensions of the simulation box were $5.87 \times 5.42 \times 12.00$ nm$^3$. All the initial ice molecules were constrained by a harmonic potential with a force constant of 1000 kJ mol$^{-1}$ nm$^{-2}$. Five independent simulations were conducted for each oligoproline with different backbone orientations parallel to the ice surface. First, the energy minimization has been conducted using the steepest descent method with a maximum force tolerance of 1000 kJ mol$^{-1}$ nm$^{-1}$. Then, the system was subsequently equilibrated at 275 K for 500 ps with NPT-MD simulation, followed by pre-equilibration at 265 K under the same ensemble. Finally, an 800 ns production NPT-MD run at 265 K was performed to monitor the ice growth inhibition activity, with restraints applied to a ~1 nm layer of water molecules below the ice slab to prevent initial ice growth in the downward direction.

To investigate the impact of P15 aggregation on ice growth, ice growth inhibition simulations were performed using four P15 chains in both aggregated and dispersed states, as obtained from metadynamics simulations. The initial simulation box dimensions were 11.75 × 11.77 × 12.00 nm³, containing approximately 46,000 water molecules and four layers of hexagonal ice.

**2.2. Metadynamics simulations.** The well-tempered metadynamics (WTMetaD),[82] a powerful and well-established enhanced sampling technique that drives the system to explore the entire free energy surface with respect to selected collective variables (CVs), was employed to investigate the conformational space of oligoproline and its aggregation property in solution. All metadynamics simulations were performed using PLUMED 2.8[83,84] plugin integrated with GROMACS 2020.7.[72]

Prior to the WTMetaD simulations, a polyproline (DP = 3, 8, 15) with PPII helix structure was solvated in water within a cubic simulation box, with dimensions at least 2 nm larger than the length of oligoproline chain. Subsequently, the system was well equilibrated in the isobaric isothermal (NPT) ensemble at 300 K. Thereafter, three different configurations of each polyproline were randomly selected from the well-equilibrated NPT-MD simulations to perform WTMetaD simulations. To probe the conformational space of oligoproline (DP=3,8,15), the radius of gyration ($R_g$) of the backbone was employed as CV to bias. After thorough validation, WTMetaD simulation parameters were set as follows: Gaussian width ($\sigma$) was 0.01 nm, and the bias factor ($\gamma$) was set to 100 for all simulations. Gaussians were deposited every 500 steps with an initial height ($W$) of 0.6, 0.8, and 1.0 kJ/mol for P3, P8, and P15, respectively. The free energy profiles were calculated using the reweighing technique proposed by Tiwary and Parrinello,[85] with final energy profiles obtained by averaging the results of three independent simulations with different initial conformations.

Additionally, the WTMetaD simulations were conducted to examine the aggregation characteristics of oligoproline with varying DPs. Initially, two oligoproline molecules (DP=3, 8, 15) were solvated in a solution containing approximately 11000 water molecules, achieving a concentration of 0.01 M (~20 mg/mL), to explore the energy landscape of oligoproline for dimer formation. In this context, the coordination number (CN) of the heavy atoms between two oligoproline molecules was used as CV to bias. Here, the CN was calculated as

$$CN = \sum_{i \in A} \sum_{j \in B} s_{ij} \qquad (1)$$

where $A$ and $B$ represent the heavy atoms of two different oligoproline molecules. A larger CN value indicates a higher aggregation propensity for oligoproline. The $s_{ij}$ is a switching function, which defined as

$$s_{ij} = \frac{1 - \left(\frac{r_{ij}}{r_0}\right)^n}{1 - \left(\frac{r_{ij}}{r_0}\right)^m} \qquad (2)$$

where $r_{ij}$ represents the distance between heavy atoms $i$ and $j$. The cutoff distance for the contact $r_0$, was set to 0.55 nm in this study. The values of n and m were set to 6 and 12, respectively, allowing for a smooth transition in the switching function. The parameters for WTMetaD simulations have been set as follows: Gaussian potentials, with $\sigma$=2.0/4.0/5.0, $\gamma$=20/20/20, and $W$=1.0/1.0/1.0 kJ/mol for P3, P8, and P15, respectively. Gaussians were deposited every 500 steps to progressively build the free energy landscape. Additionally, larger systems comprising 20 P3, 8 P8, and 4 P15 molecules were constructed to facilitate a more comprehensive investigation of the aggregation properties (multimers) of oligoproline, maintaining a similar mass concentration of approximately 40 mg/mL across all simulation systems. The CN of the heavy atoms of oligoproline was again selected as the CV to bias. In this context, equation (1) should be interpreted as a sum over all N(N−1)/2 pairs of the N heavy atoms of oligoproline. During the WTMetaD simulations, Gaussian potentials with $\sigma$=20, $\gamma$=20, and $W$=1.0 kJ/mol were deposited every 500 steps. The free energy profiles and their corresponding uncertainty were calculated using the reweighing technique of Tiwary and Parrinello,[85] which yielded relatively small uncertainty values of less than 0.5 kJ/mol in our simulations.

**2.3. Analysis Details.** The CHILL+ algorithm[86] was employed for the identification of water molecules in ice (in both the cubic and hexagonal phases) or liquid phase. To accurately describe the advancing ice front, ice molecules (identified by the CHILL+ algorithm) with fewer than three other ice molecules within 0.5 nm were also considered as liquid water. After that, interfacial water molecules located within 0.35 nm of either cubic or hexagonal ice were identified as belonging to the ice phase. To estimate the number of -$CH_2$ groups attached to the ice surface, the -$CH_2$ groups were treated as attached to the ice surface if there were six ice molecules (a hexatomic ring) within 5.5 Å (the first solvation shell) of it. The hydrogen bonds were identified with a donor-acceptor distance of less than 3.5 Å and a hydrogen-donor-acceptor angle of less than 30°.[87] Moreover, the PPII helix was identified based on the backbone dihedral angles $\phi \in$[-110,-30] and $\psi \in$[120,180].[88] The embedded depth of oligoproline on the ice surface was estimated by the $z_{PPro} - z_{ice}$, where $z_{PPro}$ represents the lowest point of the protein, and $z_{ice}$ denotes the position of ice surface along the in z-axis. All the oligoproline molecules that were bound to the ice but not overgrown by the advancing ice front in simulations were counted. The coverage area of oligoproline chains on the ice surface was determined by projecting the heavy atom positions of the polymer bound to the ice onto the xy-plane, using an atomic radius of 0.5 nm. Additionally, the interaction energies, including electrostatic and van der Waals interactions, were analyzed to assess the interactions between oligoproline and water/oligoproline during ice inhibition simulations. The Visual Molecular Dynamics (VMD) package was used for visualization purposes.[89]

## 3. RESULTS AND DISCUSSION

### 3.1. PPII helix is not the primary reason for the nonmonotonic effect in oligoproline, but the coil conformation.

It has been established that the conformation of antifreeze agents significantly influences their IRI activity.[55,56,90,91] Therefore, we first calculate the conformational free energy landscapes of oligoprolines in solution with DP = 3, 8, 15 (denoted as P3, P8, and P15, respectively) through well-tempered metadynamics simulations (see Section 2.1 for details). The radius of gyration ($R_g$) is chosen as the collective variable to bias, with a lower $R_g$ value indicating a more compact oligoproline chain. As illustrated in Figure 1a, all oligoproline chains adopt two types of energetically favorable conformations: the more compact random coil (C) and the extended linear (L) forms (labeled as PnC and PnL, with n representing the DP). A further examination of L and C conformations in P8 and P15 supports that the C conformation originates from the interspersing of the cis/trans isomers in oligoproline.[92] Furthermore, although P15 displays increased flexibility in comparison to P8, as demonstrated by its marginally lower energy barrier from L to C conformers, the free energy difference between the C and L conformations in P8 is approximately 1.5 kJ/mol lower than in P15. This indicates that the C conformation in P8 is thermodynamically more stable than in P15.

We estimate the population ratio of the C and L conformations in oligoproline across different DPs using $p(L)/p(C) = \exp((G_C - G_L)/(k_B T)) = \exp(\Delta G_{CL}/(k_B T))$, where $\Delta G_{CL}$ represents the free energy difference between two conformations, $k_B$ is Boltzmann constant, and $T$ is absolute temperature. A smaller $\Delta G_{CL}$ value results in a higher content of C conformation. From the results shown in Figure 1a, the C conformation content follows the trend: P3 > P8 > P15, with P8 exhibiting approximately twice the C conformation of P15 at the same concentration. Additionally, the two conformations of P3 show negligible differences in surface area and volume (not shown) and are therefore not differentiated in subsequent simulations.

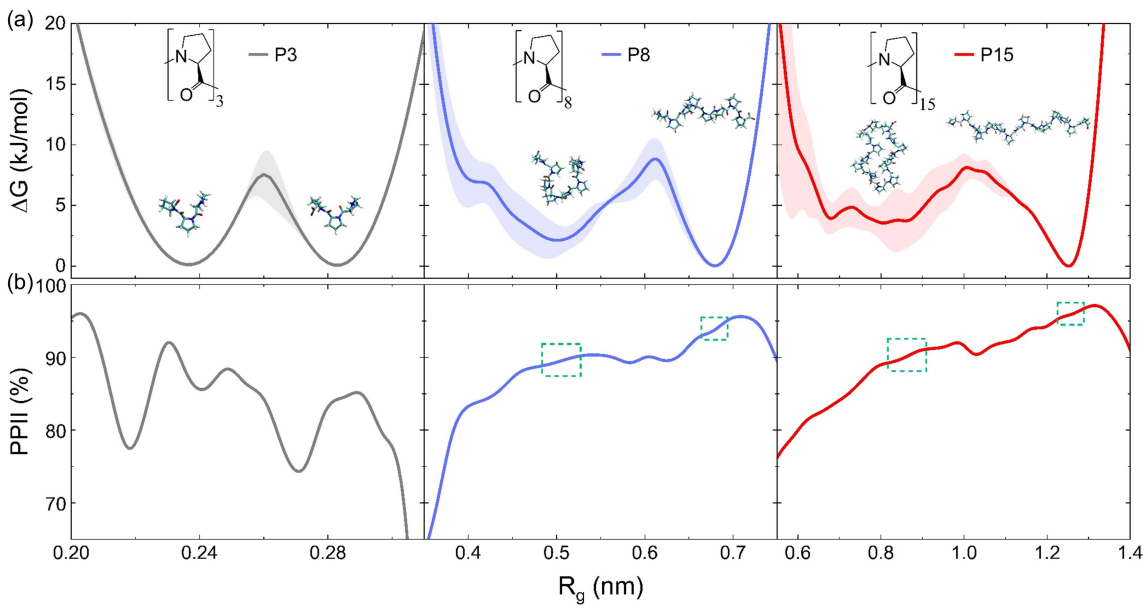

Figure 1. Conformational landscapes of oligoproline in solution at 300 K. (a) Gibbs free energy ($\Delta G$) profile as a function of the radius of gyration ($R_g$) for P3 (light gray), P8 (blue), and P15 (red). Oligoprolines are relatively rigid molecules that adopt two thermodynamically stable conformations: the linear (L) and the coil (C) forms, which are separated by energy barriers approximately 2-3 times the magnitude of thermal fluctuations at room temperature ($k_BT \approx 2.5$ kJ/mol). The shaded region represents the standard error associated with our estimate of $\Delta G$. (b) Population of the PPII helix as a function of $R_g$, colors as in (a). Dotted green squares mark the location of the $R_g$ with the thermodynamically stable conformations.

We further investigate the PPII helix content in oligoproline across different DPs. As shown in Figure 1b, both the L and C conformations in P8 and P15 are dominated by the PPII helix, with its content exceeding 88%. In contrast, P3 displays a more disordered and less defined secondary structure, with no clear correlation between PPII content and $R_g$. However, even in the case of P3, the PPII helix remains the dominant structure with a population of more than 60%. Furthermore, we calculate the total PPII content across various DPs and find that it follows the order P3 (~63%) < P8 (~80%) < P15 (~87%), indicating that the PPII helix secondary structure becomes more clearly defined with increasing DP, consistent with recent CD experimental observations.[60] Based on these results, we speculate that the dominant PPII helix structure is an intrinsic feature of oligoproline but not the primary driver of the nonmonotonic effect.

To investigate the influence of oligoproline with varying DP and conformations—specially P8 and P15 in both C and L forms, as well as P3—on ice growth, we conduct a series of independent MD simulations with the polymers being presented at the ice/water interface of the ice prismatic plane at 265 K (see Section 2.2 for details). As shown in Figure 2a, it is evident that the presence of oligoprolines significantly inhibits ice growth, with the ranking of the inhibition efficacy as follows: P8C > P15C > P8L ≈ P15L > P3. Considering the results of our free energy calculations, which indicate that P8 possesses a higher content of the C conformation compared to P15, we conclude that the overall IRI performance should follow the order P8 > P15 > P3. This conclusion aligns well with the nonmonotonic experimental results.[56]

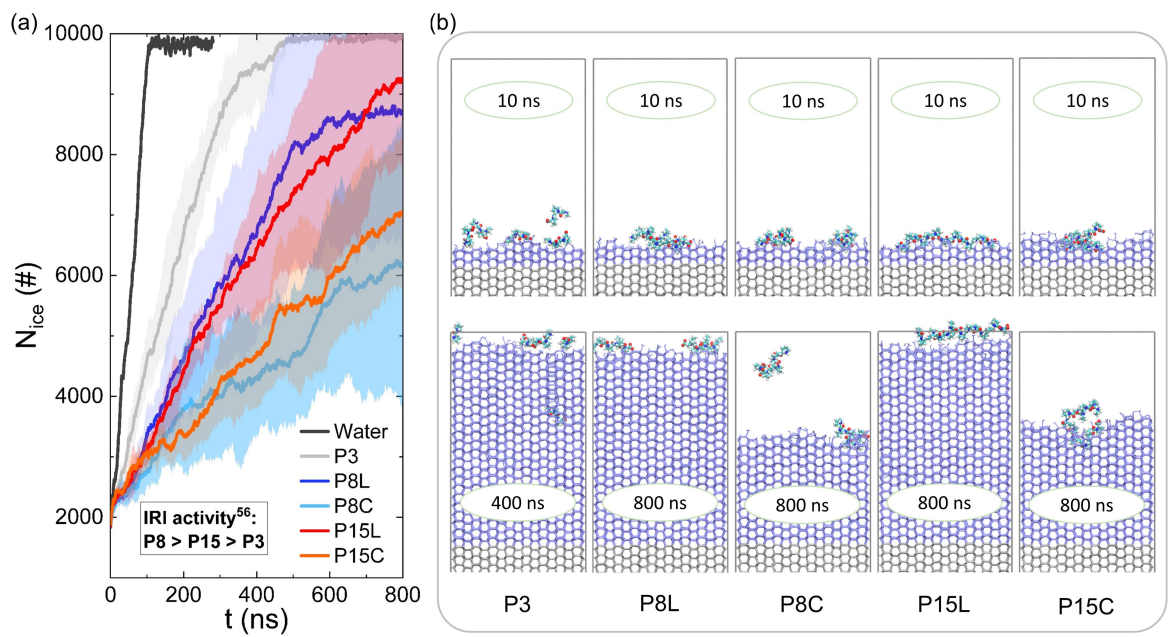

Figure 2. Ice growth inhibition ability of oligoproline with different DPs (3,8,15) and conformations (L and C). (a) Time evolution of the number of ice molecules of different simulation systems, including P3 (gray line), P8L (blue line), P8C (light blue line), P15L (red line), P15C (orange line), and control without oligoproline (black line). Data represent averages from five independent simulations, with error bars indicated as shaded regions. (b) Representative snapshots of ice growth in the presence of oligoproline.

Interestingly, we observe that oligoproline in the C conformation exhibits a significantly stronger ice growth inhibition ability compared to its L conformation (Figure 2a), even though the latter has a higher population of PPII helix structure. Therefore, we argue that the conformational preference of oligoproline, rather than the PPII content, is crucial for the nonmonotonic relationship between its IRI activity and chain length.

Figure 2b shows the representative snapshots of ice growth in the presence of oligoproline in our MD simulations. Not surprisingly, P3 is too small and tends to be easily displaced from the ice-water interface or rapidly overgrown by the advancing ice front (Figure 2 and Movie S1).

This behavior results in minimal steric hindrance to ice growth and consequently low IRI activity. In contrast, P8 and P15 in the C conformation can adhere to the ice surface for extended periods (>400 ns), with the curved regions of their chains inserted into the ice surface without being engulfed, thus demonstrating significant inhibition of ice growth (Figure 2, Movies S3,S5). However, while P15L and P8L can remain at the ice-water interface, they are less effective at inhibiting the growth of the ice front (Figure 2, Movies S2,S4), exhibiting relatively weak inhibition capabilities. Additionally, it is observed that all oligoproline chains exhibit reversible binding to ice, with the order of reversibility as follows: P8C < P15C < P8L < P15L < P3, which is consistent with their inhibition abilities. These results suggest that the adoption of the C conformation is crucial for the IRI performance of oligoproline, significantly enhancing its ice-binding capability.

**3.2. IRI activity of oligoproline: hydrophobicity, ice-binding capability, and engulfment resistance.** The ice growth/recrystallization inhibition process of antifreeze molecules involves three critical steps. First, the antifreeze molecule has to diffuse from the bulk aqueous phase to the ice-water interface and subsequently remain there.[93,94] Second, the antifreeze molecule must effectively attach to and bind tightly with the ice surface.[44] Finally, the antifreeze molecule bound to the ice surface must resist engulfment by the growing ice front.[95,96] For oligoproline, the first step is typically determined by hydrophobicity, the second step by ice-binding capability, and the last step by engulfment resistance. In the following section, we will systematically elaborate on how these factors influence the IRI performance of P3, P8, and P15.

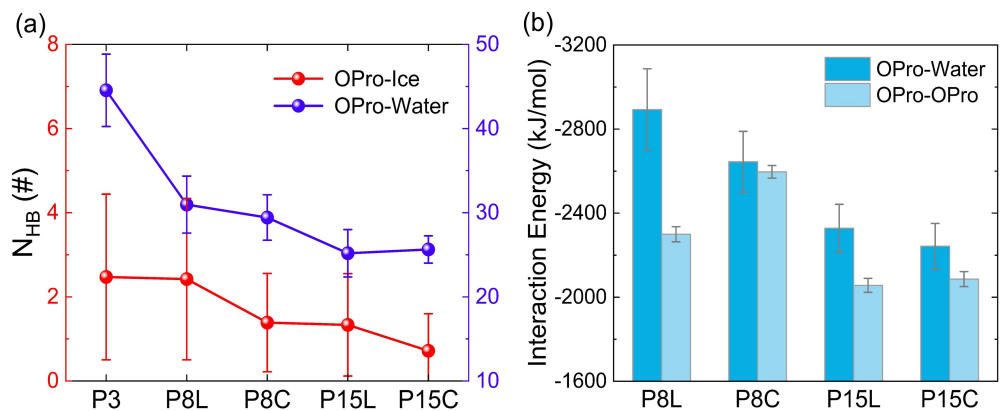

Figure 3. Interaction analysis of oligoproline-water, oligoproline-ice, and oligoproline-oligoproline. (a) Total number of hydrogen bonds formed by oligoproline (OPro) with ice (blue line) or water (red line). (b) Interaction energies between oligoproline and water (blue) or oligoproline (light blue).

In the first step, the ice surface acts as a hydrophobic wall,[88] where more hydrophobic (or less hydrophilic) oligoproline molecules are more likely to migrate to and remain at the ice-water interface. This step is the basis for the subsequent binding of oligoproline to ice. Our simulations reveal that most of the P3 chains predominantly remain in the liquid phase (Movie S1), while P8 and P15 remain continuously at the ice-water interface (Movies S2-S5). This observation indicates that the first step is particularly unfavorable for P3 due to its insufficient hydrophobicity.

In order to quantitatively assess the hydrophilicity of these oligoproline chains, we calculate the total number of hydrogen bonds formed between oligoproline molecules and water/ice in ice inhibition simulations, as shown in Figure 3a. The total number of hydrogen bonds formed between oligoproline and water follows the order P3 > P8L > P8C > P15L > P15C (Figure 3a), indicating that shorter oligoproline chains are more hydrophilic. This explains why P3 readily dissociates from the ice-water interface and dissolves in the aqueous phase, while P8 and P15 are more likely to remain at the interface. Consistent with recent studies,[59] there are almost no direct hydrogen bonds formed between oligoproline and ice. This suggests that the retention of oligoproline at the ice-water interface is not driven by enthalpic contributions from hydrogen bonding, but rather by hydrophobic entropy from the partial desolvation of the hydrophobic -$CH_2$ groups. We speculate that the hydrophobic effect of the -$CH_2$ groups facilitates the migration of oligoproline molecules to and their retention at the ice-water interface, while the hydrogen-bonding groups primarily contribute to the solubility of oligoproline.

To explore how the L and C conformations of P8 and P15 affect the first step, we calculate the interaction energies between oligoproline and oligoproline/water. As shown in Figure 3b, oligoproline in the C conformation exhibits weaker interaction energy with water compared to the L conformation, indicating that the latter is more hydrophilic. Conversely, the C conformation demonstrates greater hydrophobicity, facilitating its migration to and retention at the ice-water interface. It is also noteworthy that the C conformation exhibits a stronger intramolecular interaction than the L conformation (Figure 3b). These findings suggests that the oligoproline in the C conformation incurs a lower configurational entropy cost throughout the diffusion process and subsequent steps, thereby promoting its migration to and stabilization at the ice-water interface.

In the second step, the ice affinity of the oligoproline molecule allows it to attach to the ice surface, while robust ice-binding strength ensures firm adsorption and prevents detachment. To evaluate the ice affinity of oligoproline chains with various DP and conformations, we calculate the average number ($N_{CH_2}$) of the -$CH_2$ groups attached to the ice surface for each oligoproline molecule (Figure 4a). The results show that the $N_{CH_2}$ increases rapidly with DP and reaches a plateau around 6 when DP=8, suggesting that P8 achieves sufficient ice affinity. Notably, the $N_{CH_2}$ of P15 does not significantly exceed that of P8, indicating that only a fraction of the proline repeats in P15 effectively attach to the ice surface, even for the extended L conformation (P15L). This implies that excessively long chains, regardless of conformation, are not conducive to a perfect fit on the ice surface, since the surface of ice front is not completely flat. In contrast, the low $N_{CH_2}$ value of P3 (~2.6) reflects its weak ice affinity, attributable to the low hydrophobic dehydration energy of its -$CH_2$ groups (estimated to be less than 6 kJ/mol from previous study[88]), leading to the majority of P3 molecules detaching from the ice surface. In light of these results, we conclude that P8 demonstrates a comparable ice affinity to P15, which is sufficient for sustained adhesion to ice surfaces.

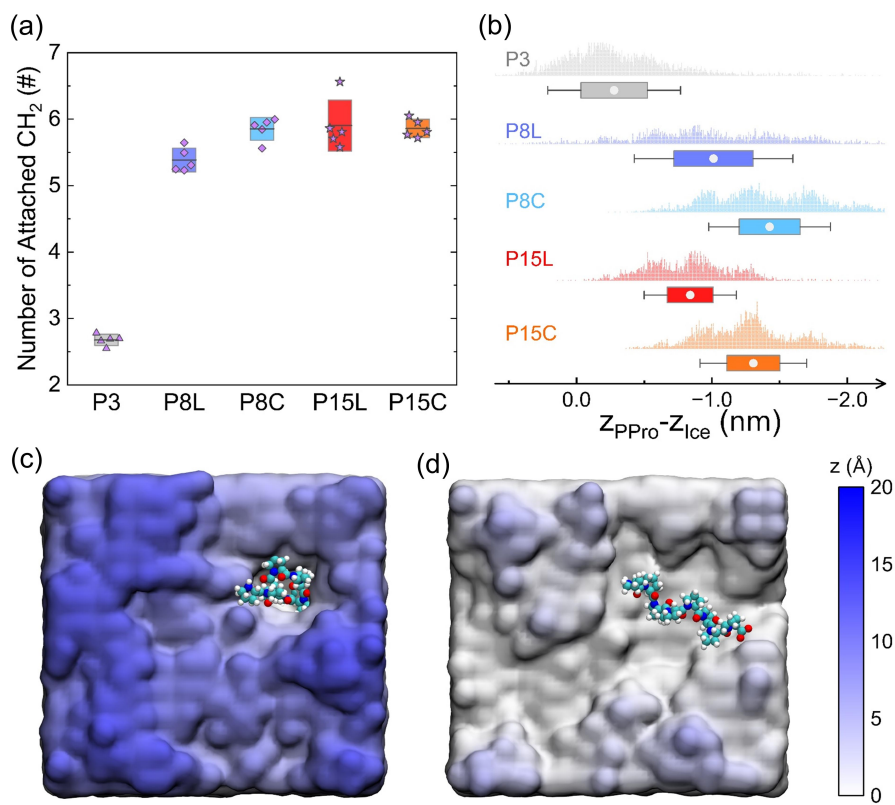

Figure 4. Binding states of oligoproline to the ice surface. (a) Average number of methylene (-$CH_2$) groups per oligoproline molecule attached to the ice surface, with data averaged over five

independent simulations (color scheme as in Figure 2a). (b) Embedded depth of oligoproline at the ice surface. (c-d) Representative binding states and configurations of P8C (c) and P8L (d), where the colors of the ice surface correspond to the height of the z-axis and the bottom atom of oligoproline is chosen as the reference point (z=0). Oligoproline molecules that stay away from the ice surface (unbound) are not shown. Additional oligoproline chains are shown in Figure S2. P8C forms the deepest trap on ice surfaces and binds most strongly to ice.

Nevertheless, high ice affinity (i.e., $N_{CH_2}$) is necessary but not sufficient for oligoproline to effectively inhibit ice growth, as demonstrated by P8L and P15L. To further analyze the ice-binding strength, we assess the embedded depth of oligoproline on the ice surface during binding but without complete engulfment. As shown in Figure 4b, more negative values of the horizontal axis indicate stronger ice-binding strength. It is observed that P8C exhibits the most significant embedded depth (Figure 4b), acting as a wedge inserted into the ice surface without being engulfed by ice, which leads to the formation of obvious folds and roughness on the ice surface (Figures 4c and S2f-j). This indicates that the ice surface adsorbed by P8C generates a larger curvature, thereby effectively inhibiting ice growth through the Gibbs-Thomson effect.

Similarly, P15C is also clearly embedded in the ice surface, as evidenced by the formation of significant folds (Figure S2e). However, P8L and P15L show only minimal embedding and produce slight surface folding (Figures 4b,d and S2b,d), which can be attributed to their continual attachment to and detachment from the ice surface. This suggests that oligoproline in the L conformation has a relatively weak ice-binding strength compared to that in the C conformation. As expected, P3 molecules exhibit only partial attachment to the ice and show minimal embedded depth, resulting in a very flat ice surface (Figure S2a,f) with limited contributions to the Gibbs-Thomson effect. In light of these results, we conclude that the superior performance of the P8 (particularly in the C conformation) compared to P15, is attributed to its greater ice-binding capability, as determined by both ice affinity and embedded depth.

In the final step, oligoproline molecule bound to ice acts like a "sand break forest", creating steric hindrance at the advancing ice front without being engulfed, thereby inhibiting ice growth. In this context, it is evident that oligoproline effectively resists engulfment by the ice front when its effective coverage area on the ice surface exceeds 3.3 nm² (Figure 5). For example, P8C and P15C are rarely engulfed during our ice growth inhibition simulation (Figure S2c,e). In contrast, when P8L and P15L attach in a parallel orientation to the advancing ice front, they are more prone to being engulfed due to their inadequate coverage area, which ultimately leads to the failure of ice growth inhibition (Figure S3). Therefore, we conclude that the effective coverage area of oligoproline molecules on the ice surface determines their ability to resist.

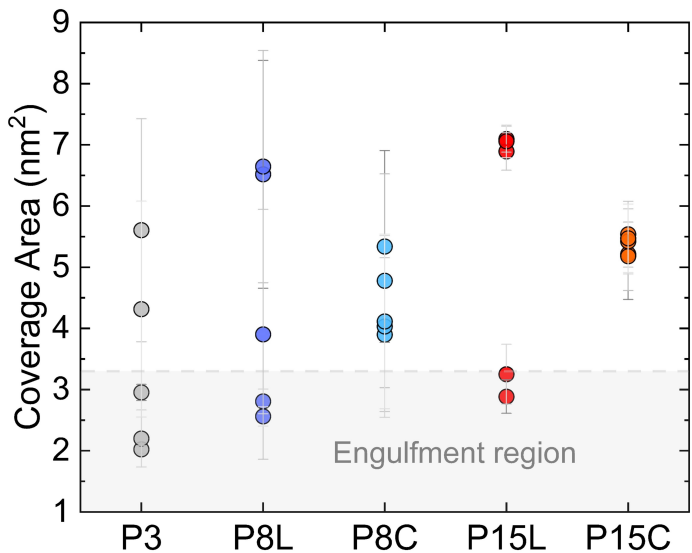

Figure 5. Average coverage area of oligoproline on the ice surface. Colors correspond to those used in Figure 2a. Data points for each oligoproline molecule are obtained from five independent simulations, with the total ice surface area of approximately 32 nm².

Unlike PVA, where the effective volume and contact area with the ice surface determine its IRI activity,[55] the origins of oligoproline's IRI activity are notably more complex. This complexity primarily stems from the fact that oligoproline interacts with ice primarily through weak hydrophobic interactions, whereas PVA relies on direct hydrogen bonding. In light of the above analyses, we conclude that the superior IRI activity of P8 is attributed to its elevated content of the C conformation. To elaborate, the enhanced hydrophobicity of P8C promotes its migration from the aqueous phase to the ice-water interface. Additionally, P8C exhibits an ice affinity comparable to that of P15 while achieving the deepest embedding depth at the ice surface, thereby strengthening its ice-binding capability. Moreover, the adequate surface coverage area of P8C effectively prevents its engulfment by the advancing ice front.

**3.3. Aggregation and its impact on the IRI performance of oligoproline.** Previous studies have indicated that aggregation can also influence the IRI activity of polymers.[61-67,69] Moreover, experimental evidence also suggests that oligoproline can form soluble aggregate at room temperature, despite the aggregation tendency becoming less pronounced as the temperature decreases.[69-71] To explore the impact of aggregation on the nonmonotonic effect in oligoproline, we then investigate the aggregation tendency and the corresponding microstructure of different oligoproline chains using WTMetaD simulations, with the coordination number (CN) as the collective variable to bias (see Section 2.2 for details). A larger CN value indicates a higher degree of aggregation.

Figure 6a represents the energy landscape for the aggregation (dimer) of oligoproline with DP=3, 8, and 15 at a low concentration (~20 mg/mL). The horizontal axis is scaled by the aggregation state ($CN_A$), with the raw data presented in Figure S4. We find that P3 and P8 show no energy minima in the aggregated state ($CN/CN_A$=1.0), while P15 exhibits a very high aggregation energy barrier (>40 kJ/mol) despite a clear minimum at the aggregated state. This suggests that these oligoproline chains tend to remain in the dispersed state rather than aggregating at this relatively low concentration.

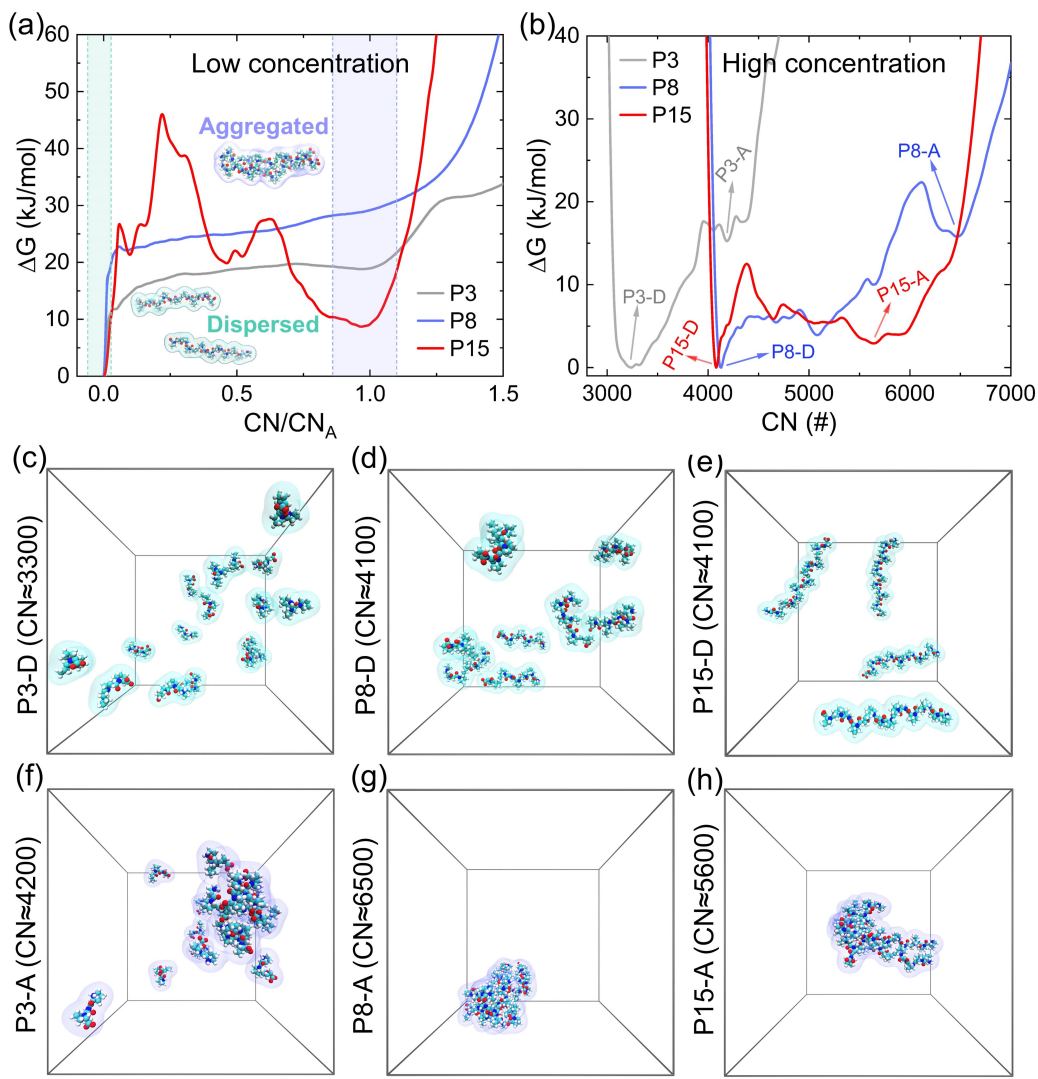

Figure 6. Aggregation tendencies of oligoprolines. (a) Free energy landscape of dimer formation for oligoprolines (DP=3,8,15) at low concentrations (~20 mg/mL), using CN as the collective variable to bias. The horizontal axis is scaled by their representative aggregation coordination number ($CN_A$) for ease of comparison, with raw data available in Figure S4. (b) Gibbs free

energy (ΔG) profile as a function of CN obtained from WTMetaD simulations at high concentrations (~40 mg/mL). The nonzero minimum CN value arises from the coordination of intramolecular adjacent atoms. (c-e) Representative dispersed configurations of P3, P8, and P15, respectively, and (f-h) representative aggregated configurations of P3, P8, and P15 from the WTMetaD simulations of (b).

We also assessed the aggregation tendency of oligoproline at a high concentration (40 mg/mL) at room temperature. As shown in Figure 6b, both dispersed and aggregated states are explored for all DPs, with representative configurations shown in Figure 5c-h. Notably, P3 and P8 not only display substantial free energies (>16 kJ/mol) in the aggregated state, but also very large aggregation energy barriers (>18 kJ/mol), indicating a strong tendency to remain in the dispersed state. In contrast, P15 shows a noticeably lower free energy of just below ~3 kJ/mol in the aggregated state, with a much smaller aggregation energy barrier (~10 kJ/mol). Therefore, we conclude that P15 exhibits a marked tendency to aggregate at high concentrations (>40 mg/mL), whereas P3 and P8 remain in the dispersed state.

Further examination of P15 aggregation reveals that it typically forms soluble, amorphous 3D structures (Figure 6f-h), as suggested by previous experiments.[69,70] Regarding the metastable states (CN≈5100 for P8 and CN≈4700 for P15), which primarily involve dimer formation (Figure S5), their impact on oligoproline's IRI activity is expected to be negligible due to their high instability and susceptibility to disruption by advancing ice front.

To investigate the impact of P15's soluble, amorphous aggregation at high concentrations on its IRI performance, we perform ice growth inhibition simulations for P15 in both aggregated and dispersed states, as shown in Figure 7. Notably, aggregated P15 exhibits a reduced ability to inhibit ice growth and is gradually overgrown by the advancing ice front. In contrast, dispersed

P15 effectively inhibits ice growth due to its enhanced coverage on the ice surface. These findings suggest that P15's disordered aggregation not only fails to enhance the ice-binding ability of oligoproline—owing to the absence of orderly arranged ice-binding sites necessary for effective interaction with the ice surface—but also limits its ability to fully cover the ice-water interface. This reduction in effective coverage area on the ice surface leads to the failure in inhibiting ice growth. Consequently, we conclude that the aggregated P15 at elevated concentrations significantly diminishes its IRI performance, thereby further amplifying the nonmonotonic effect at high concentrations, as observed in experiments.[56]

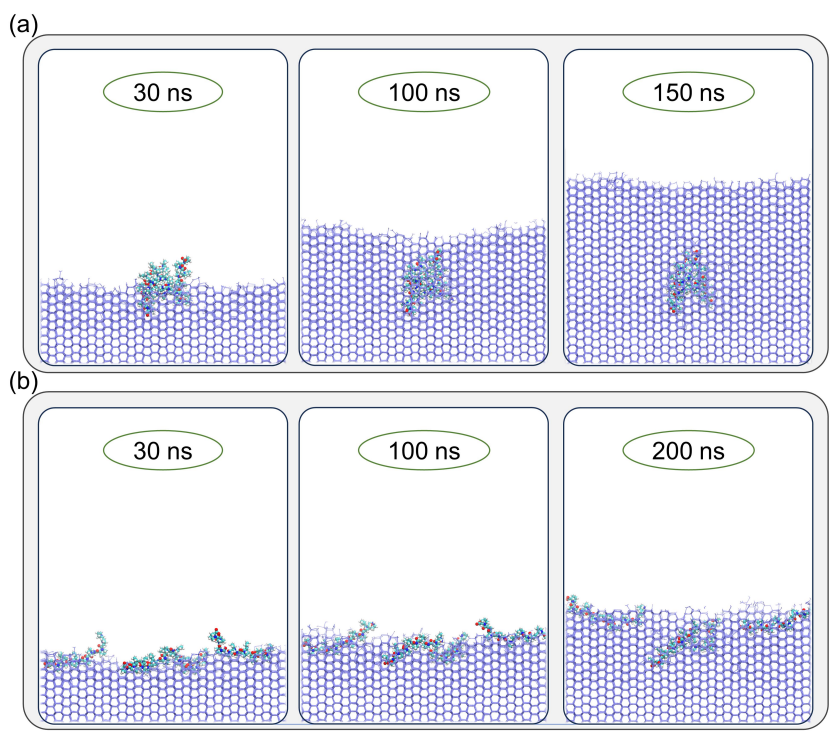

Figure 7. Impact of aggregation on ice growth inhibition of P15. (a) Aggregated P15 molecules are engulfed by the advancing ice front. (b) Dispersed P15 molecules effectively inhibit ice growth.

Furthermore, we find that there are almost no intermolecular hydrogen bonds formed in the P15 aggregations (Figure S6), indicating that hydrophobic interactions primarily drive the aggregation. As DP increases, the aggregation phenomenon becomes more pronounced due to enhanced hydrophobicity. This explains the experimental observation that PPro with DP > 100 aggregates at much lower concentrations (2 mg/mL).[69] For ultra-high DP PPro chains (DP > 100), their sufficiently strong hydrophobicity gives rise to a comparable aggregation tendency. In this context, aggregation exerts a similar influence on their IRI activity; however, these chains are of sufficient length to completely adopt the C conformation.[97] Therefore, the enhancement of IRI activity driven by the single-molecule conformation plays a more dominant role compared to the inhibitory effects induced by multi-molecular aggregation. This explains why IRI activity increases with DP beyond a certain threshold (DP>40).[69]

In addition, additional MD simulations are conducted to investigate the temperature-dependent aggregation behavior of P15. As shown in Figure S7, the aggregation tendency of P15 decreases significantly at lower temperatures (265 K), consistent with previous experimental observations.[70,71] Consequently, the impact of aggregation on its IRI activity would be further diminished at reduced temperatures due to the decreased propensity of P15 to aggregate under such conditions. These results demonstrate that the nonmonotonic effect is primarily governed by the single-molecule conformation mechanism, while the aggregation of P15 only serves to amplify this nonmonotonicity at high concentrations.

## 4. CONCLUSIONS

In this work, we employ all-atom molecular dynamics simulations to elucidate the microscopic mechanism underlying the nonmonotonic relationship between IRI activity and DP in oligoprolines. Through a systematic investigation of P3, P8, and P15, we find that the IRI activity of oligoprolines originates from the combined effect of single-molecule conformation and multi-molecule aggregation.

At low concentrations, our simulations show that the IRI-favored random coil (C) conformation content—rather than the traditionally emphasized PPII helix, whose proportion increases with DP—is the key factor driving the nonmonotonic IRI-DP behavior. P8 outperforms P15 primarily due to its higher content of the C conformation, which promotes stronger ice binding and greater resistance to engulfment by embedding more deeply into the ice surface and covering a larger interfacial area compared to the IRI-disfavored linear (L) conformation. In contrast, although P3 exhibits the highest C content among the three oligomers, its short chain length and limited hydrophobicity compromise stable and sustained interaction with ice. Furthermore, its small molecular coverage area is insufficient to prevent ice crystal growth, resulting in minimal IRI efficacy.

At high concentrations (>40 mg/mL), our simulations reveal that, in addition to the dominant role of the C conformation, the aggregation of P15 further contributes to the nonmonotonic IRI-DP relationship. Specifically, P15 tends to form soluble, amorphous aggregates at this high concentration, whereas P8 remains well-dispersed. This aggregation in P15 limits its ability to completely cover the ice surface, thereby further reducing its IRI activity.

Overall, within the context of our simulations, the content of the C conformation in oligoproline plays a decisive role in the nonmonotonic effect, while the aggregation of longer chains serves to amplify this nonmonotonicity at high concentrations. Our work not only elucidates the atomistic details of the IRI activity of oligoprolines, but also provides a valuable framework for understanding similar phenomenon in other polymers, such as zwitterionic

poly(carboxybetaine methacrylate)[57] and thymine oligomers,[58] by considering both molecular conformational preference and aggregation behavior.

## ASSOCIATED CONTENT

**Supporting Information**. Schematic representation of the simulation system used to study ice growth inhibition. Embedded states of oligoproline molecules on the ice surface. Representative snapshots of P3, P8L, and P15L being engulfed by ice. Gibbs free energy profile as a function of coordination number (CN) at low concentration (~20 mg/mL). Representative metastable configurations of P8 and P15 obtained from metadynamics simulations at higher concentration (~40 mg/mL). Total number of intermolecular hydrogen bonds formed between P15 molecules as a function of CN at high concentration (~40 mg/mL). Temperature-dependent aggregation behavior of oligoproline at 40 mg/mL. (PDF)

Movies S1-S5 describing the impact of P3, P8L, P8C, P15L, P15C on ice growth at 265 K, respectively. (MP4)

## AUTHOR INFORMATION

**Corresponding Author**

* Zhaoru Sun—School of Physical Science and Technology, ShanghaiTech University, Shanghai 201210, China.

Email: sunzhr@shanghaitech.edu.cn

**Author Contributions**

W.Y. and Z.S. designed research; W.Y. performed research; W.Y. and Y.L. analyzed data; W.Y. and Z.S. wrote the paper. All authors have given approval to the final version of the manuscript.

**Funding Sources**

This work was supported by the Double First-Class Initiative Fund of ShanghaiTech University (SYLDX0342022) and the Shanghai Rising-Star Program (23QA1406800).

**Notes**

The authors declare no competing financial interest.

**ACKNOWLEDGMENTS**

This work was supported by the Double First-Class Initiative Fund of ShanghaiTech University (SYLDX0342022) and the Shanghai Rising-Star Program (23QA1406800). The authors also thank the computing resources and technical support provided by the High-Performance Computing (HPC) Platform of ShanghaiTech University.

**ABBREVIATIONS**

PPro, Polyproline; IRI, ice recrystallization inhibition; C, random coil; L, linear.

Supporting Information

# Atomistic Insights into the Chain-Length-Dependent Antifreeze Activity of Oligoprolines

*Wentao Yang, Yucong Liao, Zhaoru Sun*[*]*

School of Physical Science and Technology, ShanghaiTech University, Shanghai 201210, China

\* Corresponding Author

E-mail address: sunzhr@shanghaitech.edu.cn (Z.S.)

## Table of Contents

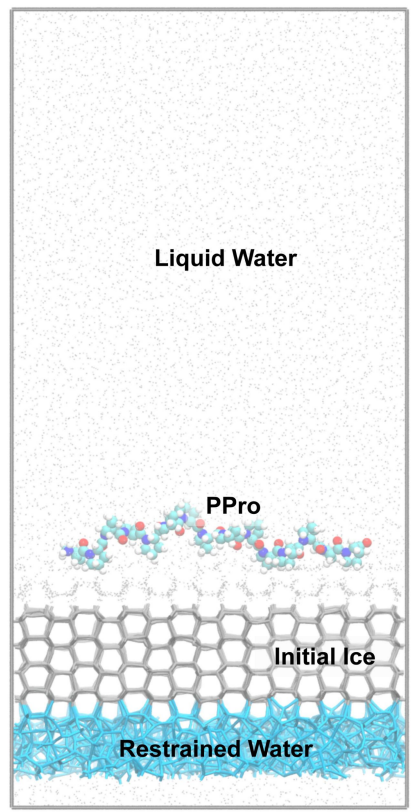

Figure S1. Schematic representation of the simulation system used to study ice growth inhibition. Four layers of initial ice (colored gray) are restrained by a harmonic potential, with the primary prismatic face oriented toward the water phases. The oxygen atoms of liquid water molecules are represented as gray points. A 1 nm thick water layer (colored cyan) below the hexagonal ice slab is restrained to prevent downward ice growth.

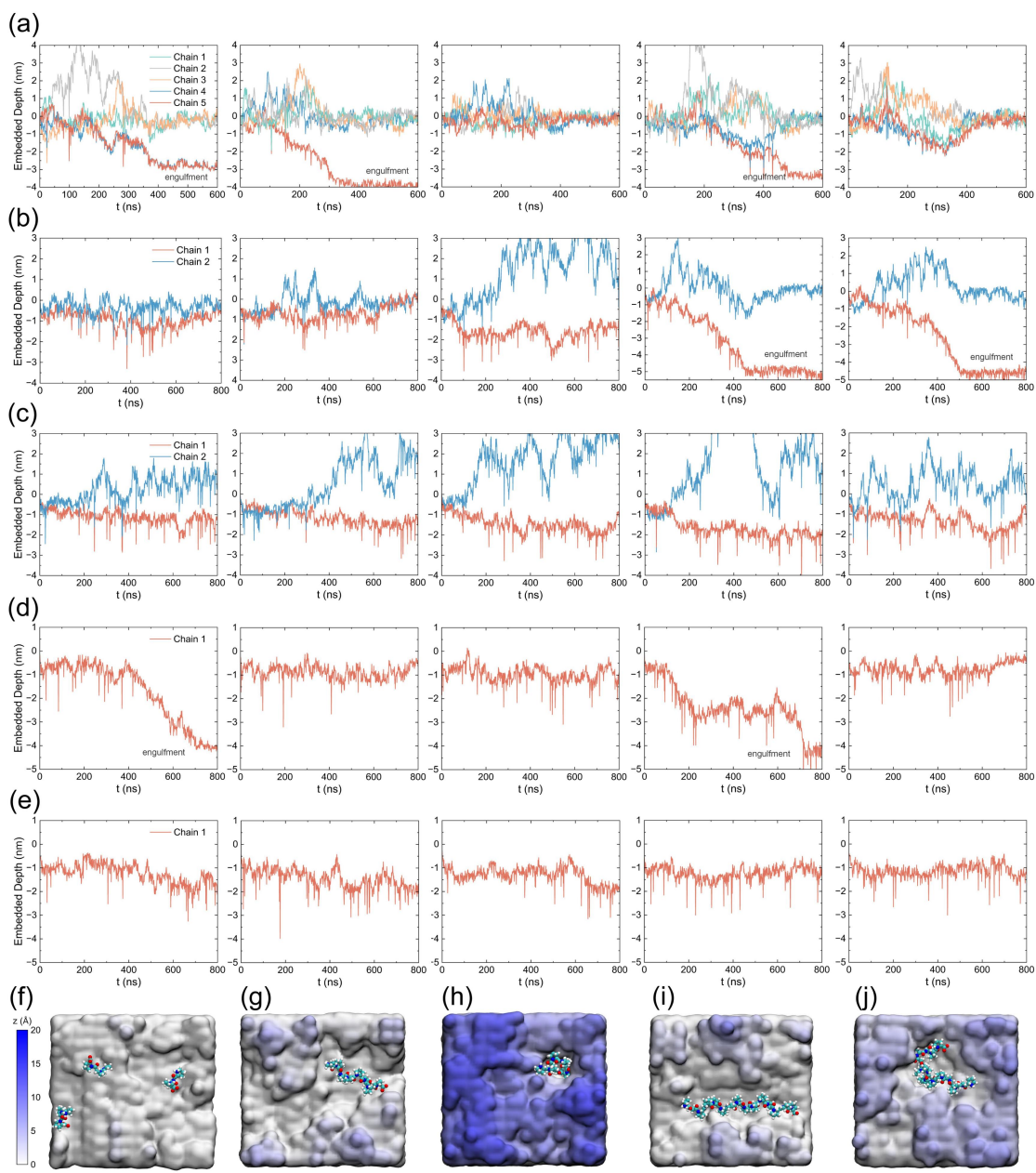

Figure S2. Embedded states of oligoproline molecules on the ice surface. (a-e) Embedded depth ($z_{PPro} - z_{ice}$) of P3 (a), P8L (b), P8C (c), P15L (d), and P15C (e) at the ice surface. (f-j) Representative binding states and configurations of P3 (f), P8L (g), P8C (h), P15L (i), and P15C (j), with the ice surface color-coded by the z-axis height. The bottom atom of each oligoproline molecule is used as the reference point (z=0). oligoproline molecules that stay away from the ice surface (unbound) are not shown.

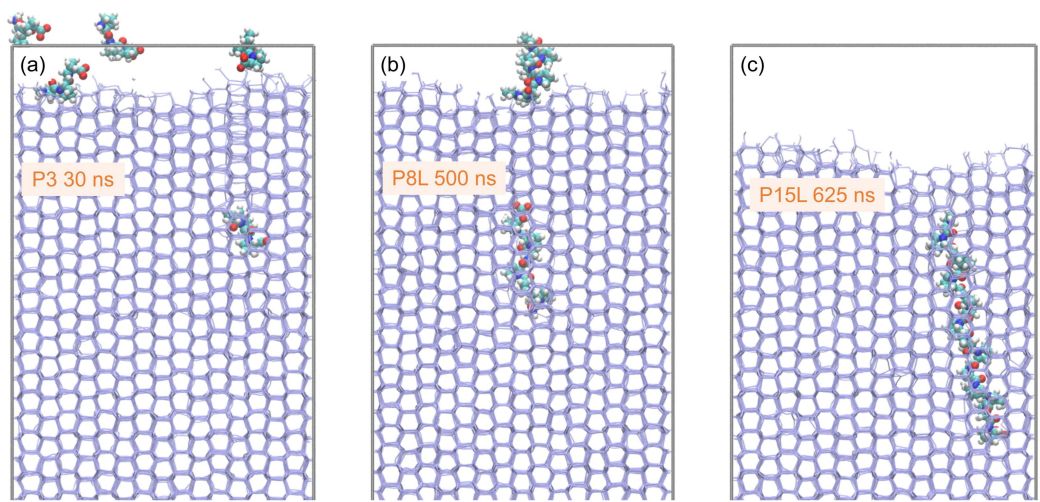

Figure S3. Representative snapshots of P3, P8L, and P15L being engulfed by ice. (a) P3 is too small and is easily engulfed by the growing ice front. (b) P8 and (c) P15 become engulfed by ice when aligned parallel to the growing ice front.

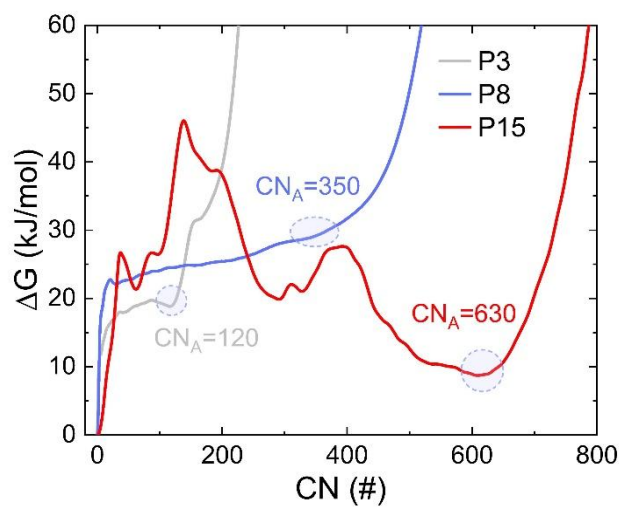

Figure S4. Gibbs free energy profile as a function of coordination number (CN) at low concentration (~20 mg/mL), with very small uncertainty values (less than 0.3 kJ/mol) obtained from our metadynamics simulations.

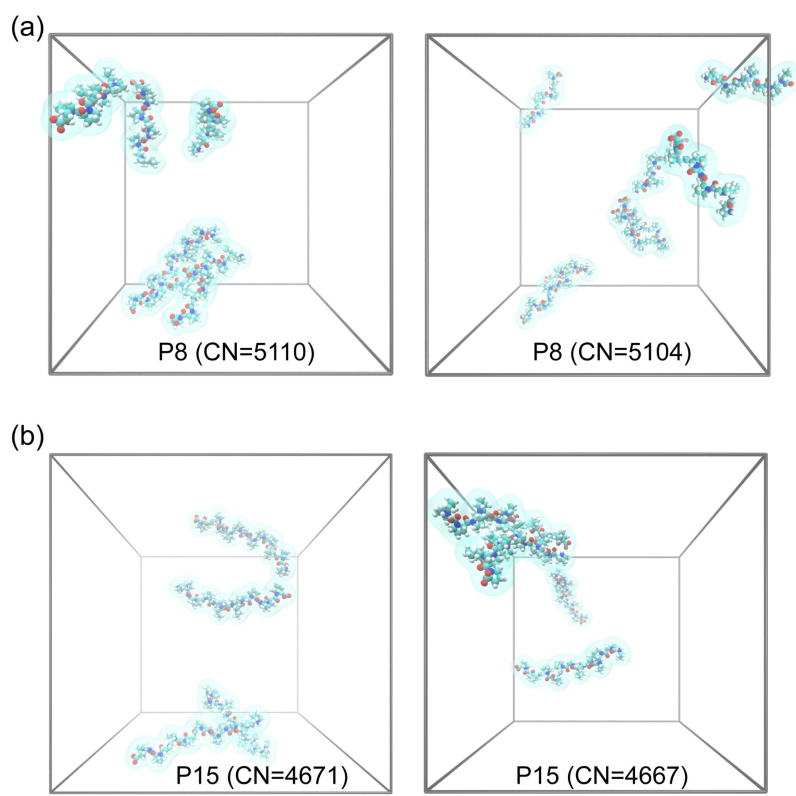

Figure S5. Representative metastable configurations of P8 (a) and P15 (b) obtained from metadynamics simulations at higher concentration (~40 mg/mL). These states primarily involve unstable dimer formations.

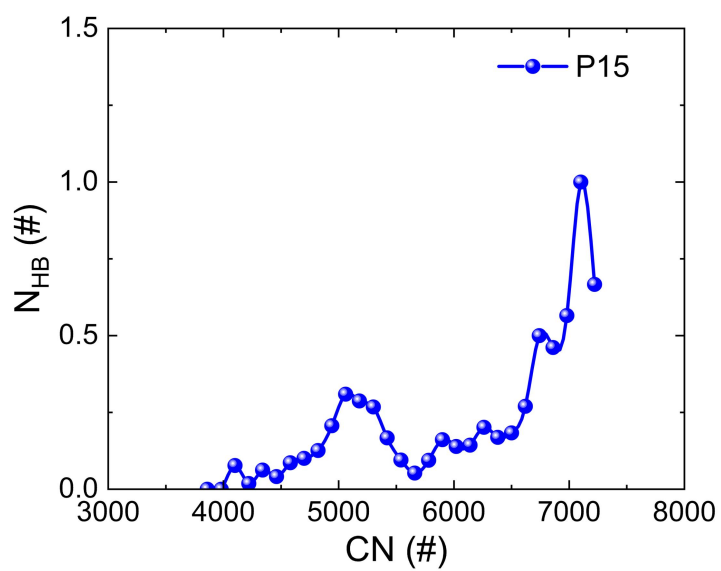

Figure S6. Total number of intermolecular hydrogen bonds formed between P15 molecules as a function of CN at high concentration (~40 mg/mL). The results show that almost no intermolecular hydrogen bonds are formed within P15 aggregations.

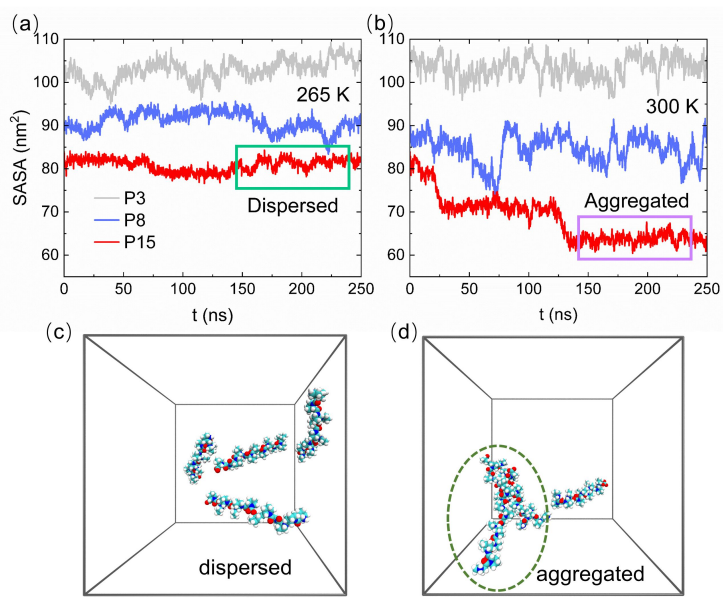

Figure S7. Temperature-dependent aggregation behavior of oligoproline at 40 mg/mL. (a) Time evolution of solvent accessible surface area (SASA) for P3, P8, and P15 at 265 K, indicating that all oligoproline chains remained in the dispersed state. (b) Time evolution of SASA for P3, P8, and P15 at 300 K. The result suggests that P15 tends to aggregated states, while P3 and P8 remain in dispersed states. (c) Representative configuration of P15 in the dispersed state (SASA ≈ 80 nm²). (d) Representative configuration of P15 in the (partially) aggregated state (SASA ≈ 65 nm²)